\documentstyle[a4wide,12pt]{article}
\begin{document}
\begin{flushright}
\baselineskip=12pt
{SUSX-TH-95/11}\\
January 1995
\end{flushright}

\begin{center}
{\huge\bf The dilaton-dominated supersymmetry breaking scenario in the
context of the non-minimal supersymmetric model \\}
\vglue 0.35cm
{G. V. KRANIOTIS \footnote {e.mail address:G.Kraniotis@sussex.ac.uk}\\}
	{\it School of Mathematical and Physical Sciences \\}
{\it University of Sussex,\\}
{\it Brighton BN1 9QH, United Kingdom \\}
\baselineskip=12pt

\vglue 0.25cm
ABSTRACT
\end{center}

{\rightskip=3pc
\leftskip=3pc
\noindent
\baselineskip=20pt
The phenomenological consequences of the dilaton-type soft supersymmetry
breaking
terms in the context of the next to minimal supersymmetric standard
model are investigated. We always find a very low top quark mass. As a
consequence such string vacua are  excluded by recent
experimental results. The viability of the solution of the $\mu$
term through the introduction of a gauge singlet field is also briefly
discussed. }

\vfill\eject
\setcounter{page}{1}
\pagestyle{plain}
\baselineskip=14pt
	The Standard Model remains unscathed in its confrontation with
experiment. Furthermore, the compatibility of the model with the recent
CDF observation \cite{CDF:gnus}, is very encouraging and enhances our belief
that we are
on the right path towards the realization of the unification program.

	However, in order to talk about the unification of all the
forces we have to see the standard model as the effective limit of a
more fundamental theory. Today the only theoretical framework which
is a strong candidate for the unification of all the interactions,
including gravity, is heterotic string theory. Supersymmetry which is
necessary in taming the radiative corrections to the Higgs boson
mass, the only still unseen particle of the standard model, is naturally
embedded in string theory.
However, supersymmetry has to be softly-broken at an energy  of
order of the electroweak scale in order to solve the hierarchy problem
mentioned above. In the
Minimal Supersymmetric extension of the Standard Model (MSSM) the
terms responsible for the breaking are customarily parametrized in terms
of four universal parameters, $M_{1/2}$, $A$, $m_{0}$, $B$, the universal
gaugino mass, the trilinear scalar terms associated with the trilinear
couplings in the superpotential, the scalar masses, and the $B$ term
associated with the higgs-doublet mixing term in the superpotential.
In string theory these parameters are in principle, predicted
but a definite answer is at present lacking due to the fact that the
supersymmetry breaking mechanism in the theory is not well understood.

	Nevertheless progress has taken place at the theoretical level,
and therefore we believe that it is an appropriate time to use  all the
present existing knowledge to study string inspired scenarios and
confront them with current experimental data. In this way
phenomenological criteria may help us select the correct string vacuum
state among the {\em plethora} of equivalent string vacua that appear at
the level of string pertubation theory, the majority of them leading to
unacceptable phenomenology.

	The process of supersymmetry breaking has to have a non-pertubative origin
since it is
well known that SUSY is preserved order by order in pertubation theory.
However, very little is known  about non-pertubative effects in string
theory, particularly in the four-dimensional case. This has led the
authors of \cite{Kap:Austin}, to parametrize the effect of SUSY-breaking by the
VEVs
of the $F$-terms
of the dilaton ($S$) and the moduli ($T_{m}$) \footnote {In string
pertubation theory, both the moduli and the dilaton are exact flat
directions of the effective potential, leaving their VEVs undetermined.
In ref. \cite{Kap:Austin} this pertubative degeneracy is assumed to be
completely lifted by the non-pertubative dynamics and VEVs for moduli
and dilaton to be induced.} chiral superfields generically
present in large classes of four-dimensional supersymmetric heterotic
strings . In a way, if supersymmetry breaking is triggered by these
fields (i.e., $\langle$$F_{S}$$\rangle$$\neq0$ or
$\langle$$F_{T}$$\rangle$$\neq0$), this would be a rather generic
prediction of string theory.

There are various possible scenarios for supersymmetry breaking which
are obtained in this model independent way. To discriminate among these
we consider a simplified expression for the scalar masses
\begin{equation}
m_{i}^{2}=m_{3/2}^{2}(1+n_{i}{\cos^{2}{\theta}})
\end{equation}
with $\tan{\theta}$$=$$\langle$$F_{S}$$\rangle$/$\langle$$F_{T}$$\rangle$
\cite{Kap:Austin}.
Here $m_{3/2}$ is the gravitino mass and the $n_{i}$ are the modular
weights of the respective matter field. There are two ways in which one
can obtain universal scalar masses, as strongly desired
phenomenologically to avoid large flavor-changing-neutral currents
(FCNCs) \cite{Ellis:Cern}: (i) setting $\theta=\pi/2$, that is
$\langle$$F_{S}$$\rangle$$\gg$$\langle$$F_{T}$$\rangle$; or (ii) in a
model where all $n_{i}$ are the same, as occurs for $Z_{2}$$\times$$Z_{2}$
orbifolds \cite{Kap:Austin} and free fermionic constructions \cite{Kal:Jul}.
In the first scenario supersymmetry breaking is triggered by  the dilaton
$F$- term and if the vanishing of the cosmological constant is imposed as a
constraint, this results in a universal scenario for the soft parameters
involved \cite{Kap:Austin}. In particular we have:
\begin{equation}
A=-M_{1/2},\; \;m_{0}=\frac{1}{\sqrt{3}}M_{1/2}
\end{equation}

	As one can see the four-dimensional soft-parameter space is
then effectively two-dimensional. In the strict-dilaton case the $B$ term
is no longer an independent parameter but is given by
\begin{equation}
B=\frac{2}{\sqrt{3}}M_{1/2}=2m_{0}
\end{equation}

	The above universal scenario leads to a natural supression of FCNC. The
dilaton dominated
scenario has been studied in the context of the minimal supersymmetric model
\cite{Bab:Phys} and in
the context of the flipped SU(5) \cite{Nan:Tex}. However, in both cases the
necessary
Higgs mixing term is provided by a bilinear mass term. The relevant term
in the superpotential is of the form
\begin{equation}
W_{\mu}={\mu}H_{1}.H_{2}.
\end{equation}
The associated soft-breaking term in the scalar potential will have the
form
\begin{equation}
BH_{1}.H_{2}
\end{equation}
It is well known that, in order to get appropriate $SU(2)_{L}$$\times$U(1)
breaking, the $\mu$ parameter has to be of the same order of magnitude as
the SUSY-breaking soft terms. This is in general unexpected since the
$\mu$-term is a supersymmetric term whereas the other soft terms are
originated after supersymmetry breaking. This is called ``the $\mu$
problem'', the reason why $\mu$ should be of the order of the soft terms.

	A very attractive solution to the problem \cite{Der:Susy} is to add an extra
singlet superfield $N$ which couples with the two Higgs doublets
superfields. The resulting superpotential will contain besides the usual
standard model terms the following term.
\begin{equation}
W_{N}={\lambda}NH_{1}.H_{2}+\frac{1}{3}kN^{3}
\end{equation}
If, the gauge singlet acquires a vacuum expectation value $\langle$$N$$\rangle$
then the
role of $\mu$ is played by $\lambda$$\langle$$N$$\rangle$ and the role of $B$
is played by
$A_{\lambda}$. The resulting model is the so called nonminimal
supersymmetric standard model. We note that such a superpotential which
contains only trilinear couplings emerges naturally in superstrings.

	It is also very interesting to point out that in general the soft
terms computed in a general class of string models are complex. The
resulting phases, are quite constrained by limits on the electric dipole
moment of the neutron (EDMN), since they give large one-loop
contributions to this CP-violating quantity. In the case of the MSSM, there are
three classes of phases associated with the parameters $A$, $M_{1/2}$, $B$ and
$\mu$
which are candidates for CP violation. These are given by
\begin{eqnarray}
\phi_{A} & = &{arg(A_{ijk}/{\lambda}_{ijk})}, \nonumber \\
\phi_{B} & = &{arg(B/{\mu})},       \nonumber \\
\phi_{C} & = &{arg(M_{a})}
\end{eqnarray}
where $M_{a}$ are the masses of the gaugino fields associated with the
three gauge group factors of the MSSM and $\lambda_{ijk}$ are the
trilinear Yukawa couplings of the chiral superfields of the theory and
$A_{ijk}$ the usual trilinear soft terms.
However, it has been shown \cite{Hal:Usa}that after a redefinition there are
only two
CP violating  phases
\begin{equation}
\phi_{A}=arg(AM^{*}_{1/2}),\;\;\;\phi_{B}=arg(BM^{*}_{1/2})
\end{equation}
These phases are constrained as we mentioned by the electric dipole
moment of the neutron. One has in fact
\begin{equation}
\phi_{A},\phi_{B}\leq10^{-3}
\end{equation}
for sparticle masses around few hundreds GeV.
The EDMN receives important contributions from both $\phi_{A}$ and
$\phi_{B}$ phases. In the dilaton-dominated scenario in the MSSM it has
been shown that the phases for $M_{a}$ and $A$ coincide and
$\phi_{A}$$\rightarrow0$ \cite{Kap:Austin}. However,  $\phi_{B}$ is in general
large and
can be sufficiently supressed only if further assumptions about the
origin of SUSY-breaking are made \cite{Kap:Austin,Choi:Phas}.

	In the case of the nonminimal model, the dilaton-dominated
scenario leads to a natural supression of the EDMN. Indeed, if the $\mu
$ problem is solved by the addition of a singlet $N$, the role of $B$ is
played by a trilinear coupling $A_{\lambda}$ whose phase is aligned with
that of the gauginos and naturally $\phi_{A}=\phi_{A_{\lambda}}=0$ .
Thus the nonminimal model
with supersymmetry breaking by the VEV of the dilaton field is a well
motivated scenario that deserves further study. However, as we shall
see, phenomenological considerations exclude the model.

    A very attractive feature of spontaneously broken effective
supergravities is the radiatively induced breaking of the electroweak
gauge symmetry \cite{Tam:Rad}.For the study of electroweak symmetry breaking we
use the one-loop
effective potential
\begin{equation}
V=V_{0}+\Delta{V_{1}}
\label{one:loop}
\end{equation}
where the radiative corrections to the scalar potential
\begin{equation}
\Delta{V_{1}}=\frac{1}{64{\pi^{2}}}{\rm Str}\left[{\cal
M}^{4}\left(\ln{\frac{{\cal M}^{2}}{Q^{2}}}-\frac{3}{2}\right)\right]
\label{rad:top}
\end{equation}
depend on the Higgs fields through the tree-level squared-mass matrix
${\cal M}^{2}$. The supertrace in equation $(\ref{rad:top})$ is given by
\begin{equation}
{\rm Str}f({\cal M}^{2})=\sum(-1)^{2J_{i}}(2J_{i}+1)f(m_{i}^{2})
\end{equation}
where $m_{i}^{2}$ denotes the field-dependent mass eigenvalue of the
{\em i}th particle of spin $J_{i}$.
In the above expression for $V$, all parameters of the theory are
running parameters, functions of the renormalization point $Q$. The use of
the one-loop effective potential guarantees the scale-independence of
the solutions \cite{Gio:REWB}.
The tree level potential $V_{0}$ contains the standard $F$- and $D$- terms, and
the
following soft-supersymmety breaking terms:
\begin{eqnarray}
(M_{1}\lambda_{1}\lambda_{1}+M_{2}\lambda_{2}\lambda_{2}+M_{3}\lambda_{3}\lambda_{3}+h_{t}A_{t}Q.H_{2}U_{R}^{c}+{\lambda}A_{\lambda}H_{1}.H_{2}N+\frac{1}{3}kA_{k}N^{3})+h.c.\nonumber\\
  +
m_{1}^{2}|H_{1}|^{2}+m_{2}^{2}|H_{2}|^{2}+m_{N}^{2}|N|^{2}+m_{Q}^{2}|Q|^{2}+m_{U^{c}}^{2}|U^{c}|^{2}+...
\end{eqnarray}
where $\lambda_{1}$, $\lambda_{2}$, and $\lambda_{3}$ denote the
gauginos of the $U(1)_{Y}$, SU(2) and SU(3) gauge groups, respectively.

	For the calculation of radiative corrections in $(\ref{rad:top})$we take into
account the corrections due to top, and stop  loops. This approximation
is correct as long as $\tan{\beta}<m_{t}/m_{b}$, where $m_{t}$ and $m_{b}$
are the top and bottom quark masses respectively and $\tan{\beta}$ is
the ratio of the VEVs of the two Higgs doublets
\begin{equation}
\tan{\beta}={\langle}H_{2}{\rangle}/{\langle}H_{1}{\rangle}
\end{equation}
It has  been shown that supersymmetry  prevents the spontaneous
breaking of CP \cite{Rom:Lett}, so that the vevs of the fields $H_{1}$, $H_{2}$
and N
are of the form
\begin{equation}
{\langle}H_{1}{\rangle}=\left(\begin{array}{c}
                                  v_{1}     \\ 0
                                 \end{array} \right),
{\langle}H_{2}{\rangle}=\left(\begin{array}{c}
                                  0      \\v_{2}
                                  \end{array} \right),
{\langle}N{\rangle}=x
\label{telos:fin}
\end{equation}
with $v_{1}$,$v_{2}$,$x$ real.

	The equations for extrema of the full scalar potential in the
directions $(\ref{telos:fin})$ in field space read
\begin{equation}
v_{1}[m_{1}^{2}+{\lambda^{2}}(v_{2}^{2}+x^{2})+\frac{1}{4}(g_{1}^{2}+g_{2}^{2})(v_{1}^{2}-v_{2}^{2})]+{\lambda}v_{2}x(kx+A_{\lambda})+\frac{1}{2}\partial{{\Delta}V_{1}}/\partial{v_{1}}=0,
\label{min:1}
\end{equation}
\begin{equation}
v_{2}[m_{2}^{2}+{\lambda^{2}}(v_{1}^{2}+x^{2})+\frac{1}{4}(g_{1}^{2}+g_{2}^{2})(v_{2}^{2}-v_{1}^{2})]+{\lambda}v_{1}x(kx+A_{\lambda})+\frac{1}{2}\partial{{\Delta}V_{1}}/\partial{v_{2}}=0,
\label{min:2}
\end{equation}
\begin{equation}
x[m_{N}^{2}+{\lambda^{2}}(v_{1}^{2}+v_{2}^{2})+2k^{2}x^{2}+2{\lambda}kv_{1}v_{2}+kA_{k}x]+{\lambda}A_{\lambda}v_{1}v_{2}+\frac{1}{2}\partial{{\Delta}V_{1}}/\partial{x}=0
\label{min:3}
\end{equation}

	We now describe our numerical procedure. At the string
unification scale the following relations hold
\begin{eqnarray}
g_{1}=g_{2}=g_{3}{\equiv}g_{U} \nonumber \\
M_{1}=M_{2}=M_{3}{\equiv}M_{1/2}  \nonumber \\
h_{t}=h^{t}_{U},\lambda=\lambda_{U},k=k_{U} \\
m_{i}^{2}{\equiv}m_{0}^{2}=\frac{1}{3}M_{1/2}^{2},&i=1,2,N,Q,U^{c} \nonumber\\
A_{t}=A_{\lambda}=A_{k}{\equiv}A=-M_{1/2}\nonumber
\end{eqnarray}

We scan the parameter space of the model as follows:First we choose a set of
initial values for the parameters,$h^{t}_{U}$, ${\lambda_{U}}$, $k_{U}$,
$M_{1/2}$ at the scale of O$(10^{16})$GeV. Then by using the well known set
of RGE at the one-loop order we evolve the relevant parameters down to the
electroweak scale
of O(100)GeV and insert their numerical values into the minimization
equations $(\ref{min:1})-(\ref{min:3})$. The latter are highly non-linear
algebric equations
with independent variables the three vacuum expectation values.
In order to solve them we use numerical routines from the NAG library
which iteratively converge to a solution. We start with an initial guess for
the three
vacuum expectation values and the routines employed quickly converge to
the solution for $v_{1}, v_{2}, x$. We seek solutions where three nonzero
vacuum
expectation values develop. Our solutions are nontrivially constrained,
by the physical requirement that the correct mass for the $Z^{0}$
boson  must be reproduced. Furthermore, in order to
guarantee a global minimum, we demand that all the physical Higgs bosons
have positive mass squared and that the vacuum expectation value of the
Higgs potential is negative and therefore energetically preferable over
the symmetric minimum. An essential check of the correctness of the
minimization of the one-loop effective potential is the appearance of
the charged and neutral massless Goldstone bosons, which indicate that
the charged and neutral Higgs field dependence has been included
properly in the ${\cal M}$ matrices appearing in $(\ref{one:loop})$. The
running top quark mass obtained in our
procedure is related to the experimentally observable pole mass by
\cite{Top:Fer}
\begin{equation}
m^{pole}_{t}=m_{t}(m_{t})\left[1+\frac{4}{3}\frac{\alpha_{s}(m_{t})}{\pi}+K_{t}\left(\frac{\alpha_{s}(m_{t})}{\pi}\right)^{2}\right]
\label{top:pole}
\end{equation}
where
\begin{equation}
K_{t}=16.11-1.04\sum_{i=1}^{5}\left(1-\frac{M_{i}}{M_{t}}\right)
\end{equation}
where $M_{i}$, i=1,...,5, represent the masses of the five lighter
quarks. We solve $(\ref{top:pole})$ by using the nonlinear equation
solution routines described above for the minimization of the effective
potential.

	The experimental constraints that we impose are summarized in
table \ref{exper}.
\begin{table}[h]
\begin{center}
\begin{tabular}{|c|c|}                        \hline
Particle        & Experimental Limit (GeV) \\ \hline \hline
 gluino         &           120            \\ \hline
squark,slepton  &            45            \\ \hline
chargino        &            45            \\ \hline
neutralino      &            20            \\ \hline
light higgs     &            60            \\ \hline
\end{tabular}
\end{center}
\caption{Experimental constraints}
\label{exper}
\end{table}
We also require that $m^{2}_{\tilde{\nu}}>0$ to avoid a ${\Delta}L{\neq}0$
vacuum \cite{Der:Susy} and that $m_{t}^{pole}> 131GeV$ \cite{D0:abach}.

\newpage

As it turned out the model failed to satisfy all the constraints imposed from
radiative electroweak breaking and the experiment. The
results of our research are summarized in table 2. In particular we list
values of the relevant parameters which reproduce the correct $Z^{0}$
boson mass, $m^{2}_{\tilde{\nu}}>0$ and gluino masses above $120GeV$. The
single entry for $m_{t}^{pole}$ in table 2 indicates the maximum value
of the physical top quark mass obtained in our study. For our notation
in table 2 see Ellis et al. in ref. \cite{Der:Susy}.

\begin{table}
\begin{center}
\begin{tabular}{||l|c|c|c||} \hline\hline
Parameter     		& $\left| k_{U}\right|=0.01$  & $\left|k_{U}\right|=0.1$
&$\left|k_{U}\right|=1.0$ \\ \hline
$h_{U}^{t}$   		& 0.16-0.182    & 0.16-0.184   &0.16-0.189  \\
$\lambda_{U}$ 		& 0.16-0.35     & 0.19-0.35    &0.34-0.57   \\
$h_{t}$       		& 0.47-0.52     & 0.47-0.53    &0.46-0.54   \\
$\lambda$     		& 0.2-0.403     & 0.24-0.405   &0.34-0.5    \\
$\left|k\right|$        & 0.01          & 0.082-0.093  &0.43-0.49   \\

$m_{t}$(GeV)            &71-91          & 71 -90       &70-91       \\
$m_{t}^{pole}$          &97             & 96           &96          \\
$\tan{\beta}$           &1.42-7.6       &1.41-4.51     &1.4-3.1     \\
    r                   &0.09-0.52      &0.21-0.52     &0.34-0.5    \\

$m_{\tilde{g}}$(GeV)    &120-192        & 121-200      &120-218     \\
$m_{\tilde{\nu}}$(GeV)  &2.3-37         & 3.12-39      &9.2-51      \\
$m_{\tilde{e_{R}}}$(GeV)&40-68          & 41-66        &41-69       \\
$m_{\chi_{1}^{+}}$(GeV) &12-67          & 28-68        &43-71       \\
$m_{\chi_{1}^{0}}$(GeV) &23-35          & 23-38        &9-39        \\

$m_{H^{\pm}}$(GeV)      &77-89          &78-89         &82-98         \\
$m_{P_{1}}$(GeV)        &6.6-11         &29-36         &62-103        \\
$m_{P_{2}}$(GeV)        &62-108         &74-104        &104-114       \\
$m_{S_{1}}$(GeV)        &10-39          &21-39         &12-44         \\
$m_{S_{2}}$(GeV)        &49-92          &59-88         &89-97         \\
$m_{S_{3}}$(GeV)        &95-108         &95-108        &108-134       \\
\hline\hline
\end{tabular}
\end{center}
\caption{Typical parameter values that emerge from the renormalization
group analysis which are consistent with correct electroweak breaking}
\label{analysis}
\end{table}
	As one can see from the solutions of the renormalization group
the values of the trilinear Yukawa couplings of the model that give
correct electroweak breaking are very constrained. In particular the top
Yukawa coupling is always small implying a very low top quark mass in
conflict with published experimental results \cite{D0:abach}. Higher
values of the top Yukawa coupling require very small gluino masses in order
that the constraint of correct electroweak breaking is satisfied.
Also higher values for $h_{U}^{t}$ tend to generate VEVs for the
sneutrino field. We do not list in table 2 lower values than 0.16 for the
$h_{U}^{t}$ coupling.The
lightest Higgs mass eigenstate, $m_{S_{1}}$, is always below the threshold of
60 GeV. Also,
from the study of the vacuum expectation value of the gauge singlet
Higgs field we conclude that the generated $\mu$ term is only a few GeV,
implying therefore a light chargino mass,$m_{\chi_{1}^{+}}$,  sometimes below
the current experimental limit.
At this point we must mention that it is possible to generate a
different set of VEVs, which give $m_{t}{\sim}130 GeV$, by starting with higher
values for the top and $\lambda$ Yukawas. The feature of the latter set is the
large vacuum
expectation value of the gauge singlet of O(1)TeV. However, this set of
VEVs lead to an unstable vacuum since $\langle$$V_{Higgs}(1-loop)$$\rangle$$>0$
and the Higgs mass squared eigenvalues are not all positive. This is in
agreement with the work in ref.\cite{Sav:Par} where an upper bound for
the $\lambda_{U}$ Yukawa has been obtained, $\lambda_{U}<0.55$. In our
case the values of $\lambda_{U}$ which generates the large VEV for the
gauge singlet are such that $\lambda_{U}{\geq}0.65$ for
$h_{U}^{t}{\geq}0.35$. All the above facts make the nonminimal model with
dilaton-dominated
soft-terms excluded by experiment. It is worthwhile to
mention that the no-scale scenario with the supersymmetry breaking
driven by the $F$-terms of the moduli fields \footnote{ The
dilaton-dominated scenario is a no-scale scenario driven by the $F$-terms
of the dilaton superfield.}is also not a viable scenario
in the context of supersymmetric models with a gauge singlet
\cite{Sav:Par}.
Thus the supersymmetric models with a gauge singlet Higgs field seem difficult
to
reconcile with universal no-scale scenarios. We remind the reader that
no-scale supergravity is the infrared limit of superstring theory
\cite{Wit:Pri}. Alternative mechanisms for the generation of the $\mu$
term exist in the context of string theory \cite{Kap:Austin}. From the
phenomenological point of view string inspired univeral no-scale supergravity
models with two Higgs doublets are preferable to models with the
extra gauge singlet Higgs field. However, the latter with more general
soft-supersymmetry breaking terms can lead to acceptable phenomenology
though is difficult to have definite predictions for the particle
spectrum due to the high-dimensionality of the parameter space \cite{Sav:Par}.
On the other hand, if the problem of the $\mu$ term in string theory is solved
by
the introduction of a gauge singlet superfield this might lead to a
departure from the universality of the soft SUSY breaking terms. \footnote{ Non
universal
 scenarios for the soft-terms emerge naturally on orbifold constructions
\cite{Kap:Austin}.}.

\section*{Acknowledgments}
I wish to thank my supervisor Professor D. Bailin for useful suggestions and
for
reading  the manuscript.


\newpage
\begin{thebibliography}{99}
\bibitem{CDF:gnus} F. Abe et al.: Phys. Rev. D50, 2966(1994)
\bibitem{D0:abach} S. Abachi et al.: Phys. Rev. Lett. 72, 2138(1994)
\bibitem{Kap:Austin} V. Kaplunovsky and J. Louis:
Phys.Lett. B 306 (1993) 269;A. Brignole, L.E. Ib$\acute{a}$$\tilde{n}$ez, and
C.Munoz: Nucl.
Phys. B422 (1994)125
\bibitem{Ellis:Cern} J. Ellis and D. V. Nanopoulos: Phys. Lett. B
110(1982) 44
\bibitem{Kal:Jul} S. Kalara, J.L. Lopez, and D.V. Nanopoulos: Phys.
Lett. B 269 (1991) 84
\bibitem{Hal:Usa} M. Dugan, B. Grinstein and L. Hall: Nucl. Phys.
B255(1985) 413
\bibitem{Choi:Phas} K. Choi: Phys. Rev. Lett. 72 (1994) 1592
\bibitem{Bab:Phys} R. Barbieri, J. Louis, and M. Moretti: Phys. Lett. B
312 (1993) 451
\bibitem{Nan:Tex} J. Lopez, D. Nanopoulos, and A. Zichichi: Phys. Lett.
B 319 (1993) 451
\bibitem{Der:Susy} H.P. Nilles, M. Srednichi and D. Wyler: Phys. Lett.
B120 (1983) 346;
J.M. Frere, D.R.T. Jones and S. Raby: Nucl.Phys. B222(1983)11;
J.P. Derendinger and C. Savoy: Nucl. Phys. B237(1984)307;
L.E. Ib$\acute{a}$$\tilde{n}$ez and J. Mas: Nucl. Phys. B286 (1987)107;
J. Ellis, J.F. Gunion, H.E. Haber, L. Roszkowski and F. Zwirner:
Phys.Rev.D39(1989)844;
M. Drees: Inter. J. Mod. Phys. A4 (1989) 3635
\bibitem{Rom:Lett} J. Romao: Phys Lett. B 173 (1986) 309
\bibitem{Gio:REWB} G. Gamberini, G. Ridolfi, and F. Zwirner: Nucl. Phys.
B331 (1990) 331
\bibitem{Tam:Rad} L. Ibanez, G.G. Ross: Phys. Lett. B110 (1982) 215;
L. Alvarez-Gaume, J. Polchinsky and M. Wise: Nucl. Phys. B221 (1983)
495; J. Ellis, J.S. Hagelin, D.V. Nanopoulos and K. Tamvakis: Phys.
Lett. B125 (1983) 275; M. Claudson, L. Hall and I. Hinchliffe: Nucl.
Phys. B228 (1983) 501; C. Kounnas, A.B. Lahanas, D.V. Nanopoulos and M.
Quiros: Nucl. Phys. B236 (1984) 438; L.E. Ib$\acute{a}$$\tilde{n}$ez and C.
Lopez: Nucl.
Phys. B233 (1984) 511; A. Bouquet, J. Kaplan, C.A. Savoy: Nucl. Phys.
B262 (1985) 299
\bibitem{Top:Fer} N. Gray, D. Broadhurst, W. Grafe, and K. Schilcher:
Z. Phys. C48 (1990) 673
\bibitem{Sav:Par} U. Ellwanger, M.R. de Traubenberg, C.A. Savoy: Phys.
Lett. B 315(1993) 331
\bibitem{Wit:Pri} E. Witten: Phys. Lett. B155 (1985) 151
\end{thebibliography}
\end{document}